\documentclass[cits]{PoS}

\RequirePackage{xspace}
\def\to                 {\ensuremath{\rightarrow}\xspace}

\def\piz   {\ensuremath{\pi^0}\xspace}
\def\pip   {\ensuremath{\pi^+}\xspace}
\def\pim   {\ensuremath{\pi^-}\xspace}

\def\Kaon  {\ensuremath{K}\xspace}
\def\Kbar  {\kern 0.2em\overline{\kern -0.2em K}{}\xspace}

\def\Kz    {\ensuremath{K^0}\xspace}
\def\Kzb   {\ensuremath{\Kbar^0}\xspace}
\def\KzKzb {\ensuremath{\Kz \kern -0.16em \Kzb}\xspace}
\def\Kp    {\ensuremath{K^+}\xspace}
\def\Km    {\ensuremath{K^-}\xspace}
\def\Kpm   {\ensuremath{K^\pm}\xspace}

\def\KpKm  {\ensuremath{\Kp \kern -0.16em \Km}\xspace}
\def\KS    {\ensuremath{K^0_{\scriptscriptstyle S}}\xspace}
\def\KL    {\ensuremath{K^0_{\scriptscriptstyle L}}\xspace}
\def\Kbar  {\kern 0.2em\overline{\kern -0.2em K}{}\xspace}

\def\Dbar    {\kern 0.2em\overline{\kern -0.2em D}{}\xspace}

\def\Dz      {\ensuremath{D^0}\xspace}
\def\D      {\ensuremath{D}\xspace}
\def\Dzb     {\ensuremath{\Dbar^0}\xspace}
\def\DzDzb   {\ensuremath{\Dz {\kern -0.16em \Dzb}}\xspace}
\def\Dp      {\ensuremath{D^+}\xspace}
\def\Dm      {\ensuremath{D^-}\xspace}

\def\DpDm    {\ensuremath{\Dp {\kern -0.16em \Dm}}\xspace}

\def\Bbar    {\kern 0.18em\overline{\kern -0.18em B}{}\xspace}

\def\Bpm     {\ensuremath{B^\pm}\xspace}

\title{Quantum-correlated \D Decays at CLEO-c}

\ShortTitle{Quantum-correlated \D Decays at CLEO-c}

\author{\speaker{Stefania RICCIARDI}%
         \thanks{On behalf of the CLEO-c Collaboration}\\
        STFC Rutherford Appleton Laboratory\\
        E-mail: \email{stefania.ricciardi@stfc.ac.uk}}

\abstract{

The 818 fb$^{-1}$ dataset collected at the $\psi$(3770) resonance in the CLEO-c
detector offers unique possibilities for measuring strong-phase
differences in neutral \D decays. We report results for D
decays to $\KS\pip\pim$ and $\KS\Kp\Km$. The measurements require that both \D
mesons in the event are fully reconstructed, with one decaying to
the signal mode of interest, and the other to a $CP$ eigenstate, or a flavour-specific state, or $K^0_{S,L} h^+ h^-$ ($h = \pi$ or \Kaon). The strong-phase differences extracted from these decays are important inputs to
the determination of the CKM angle $\gamma$ with $\Bpm\to D(\KS h^+h^-)\Kpm$ decays. 
}

\FullConference{European Physical Society Europhysics Conference on High Energy Physics\\
                 July 16-22, 2009\\
                 Krakow, Poland}

\begin{document}

\section{Introduction}
Quantum-correlation in the coherent $\psi(3770)\to \Dz\Dzb$ decay provides direct sensitivity to the relative strong-phase difference $\Delta\delta_D$ between $\Dz$ and $\Dzb$ decaying to a common final state $f_D$.
Here, we report on the first determination of $\Delta\delta_D$ for $D$ decays\footnote{Here, and in the following, $D$ denotes either \Dz or \Dzb.} 
to $\KS\pip\pim$~\cite{qing} and a preliminary result for $D\to\KS\Kp\Km$. These results will play a crucial role in the measurement of the CKM angle $\gamma$ from $\Bpm\to D(\KS h^+ h^-)\Kpm$ ($h=\pi$ or $\Kaon$) with a model-independent approach~\cite{GirietAl,Bondar}.
Both results are obtained by CLEO-c with the full 818 pb$^{-1}$ dataset of $e^+e^-$ collisions at the $\psi(3770)$ resonance. The clean environment and the excellent performance of the detector allow CLEO-c to reconstruct both the signal and the recoiling $D$ meson ({\em D-tag}) with high efficiency and purity.

\section{Binned measurements of $\Delta\delta_D$ for $D\rightarrow\Kz\pip\pim$ and $D\rightarrow\Kz\Kp\Km$}

Measurements of $\Delta\delta_D$ are performed separately for $D\to\Kz\pip\pim$ and $D\to\Kz\Kp\Km$ in 2N bins of the ($m_+$, $m_-$) Dalitz plot, where $m_\pm \equiv m^2(K^0 h^\pm)$. The Dalitz plane is  symmetrically divided about the diagonal ($m_+$ = $m_-$), with N = 8 for $K^0\pip\pim$ and N = 3 for  $K^0\Kp\Km$. Bins of equal size in $\Delta\delta_D$, according to the BaBar isobar models for $D\to\KS\pip\pim$~\cite{isobarBaBar} and
$D\to\KS\Kp\Km$~\cite{BaBar}, are chosen to minimise the strong-phase difference variations within a bin. This choice gives increased sensitivity to $\gamma$ compared to rectangular bins, %in $(m_+, m_-)$
without introducing any model uncertainty~\cite{Bondar}.

The amplitude-weighted mean cosine, $c_i$, and sine, $s_i$, of $\Delta\delta_D$ in each bin\footnote{Bins below the symmetry axis are indexed with $i$, and those above with $-i$ ($i = 1, N$).} are defined by
$$ c_i = \frac{a_D^2}{\sqrt{K_i K_{-i}}} \int_i |f_D(m_+, m_-)||f_D(m_-, m_+)| cos[\Delta\delta_D(m_+, m_-)] dm_+dm_- ,$$
and
$$ s_i = \frac{a_D^2}{\sqrt{K_i K_{-i}}} \int_i |f_D(m_+, m_-)||f_D(m_-, m_+)| sin[\Delta\delta_D(m_+, m_-)] dm_+dm_- ,$$
where $a_D$ is a normalisation factor and $K_i$ is the number of events in the $i^{th}$ bin of the flavour-tagged $\KS h^+ h^-$ Dalitz plot.
The $c_i$ coefficients can be determined from the event yields of CP-tagged \D decays. Mixed CP-tagged events, such as $D\to\KS\pip\pim$ vs $D\to\KS\pip\pim$, are sensitive to both $c_i$ and $s_i$.
Analogous quantities ($c_i\sp{\prime}$, $s_i\sp{\prime}$, and $K_i\sp{\prime}$) can be defined for $D\to K^0_L h^+ h^-$. These primed quantities can exhibit small differences from the corresponding un-primed ones due to additional doubly-Cabibbo suppressed contributions in the $D\rightarrow K^0_L h^+ h^-$ amplitude.

The coefficients $c_i$, $s_i$, $c_i\sp{\prime}$, and $s_i\sp{\prime}$ are simultaneously extracted, with a maximum likelihood fit, from the background-subtracted and efficiency-corrected yields for CP-tagged, $\KS h^+ h^-$-tagged and flavour-tagged $D\to K^0_{S,L}\pip\pim$ (or $D\to K^0_{S,L}\Kp\Km$) decays. 

%A constraint on the differences between primed and unprimed coefficients 
%from the model prediction is used in the fit.

\section{Data samples} 
A total of approximately 23,000 events are selected for the $D\to K^0_{S,L}\pip\pim$ analysis and 1,900 for the $D\to K^0_{S,L}\Kp\Km$ one, adding all the $D$-tags listed in Table~\ref{tab:tag1} and~\ref{tab:tag2}. The addition of $D\to\KL h^+ h^-$ decays more than doubles the useful data samples. %The missing-mass is used to infer the presence of a \KL.% in the $D$ decay. %which escapes detection. 

\begin{table}
\begin{center}
{\scriptsize{
\begin{tabular}{ll}
\hline
Tag Group & Opposite-side Tags \\ \hline
$K^0_S\pip\pim$ vs CP-even   & $\Kp\Km$, $\pip\pim$, $K^0_S\piz\piz$, $K^0_L\piz$ \\ 
$K^0_S\pip\pim$ vs CP-odd    & $K^0_S\piz$, $K^0_S\eta(\gamma\gamma)$, $K^0_S\omega(\pip\pim\piz)$\\
$K^0_S\pip\pim$ vs $K^0\pip\pim$ & $K^0_S\pip\pim$, $K^0_L\pip\pim$\\      
$K^0_S\pip\pim$ vs Flavour & $\Kp\pim$, $\Kp\pim\piz$, $\Kp\pim\pip\pim$, $\Kp e^- \nu_e$ \\ 
$K^0_L\pip\pim$ vs CP-even & $\Kp\Km$, $\pip\pim$ \\ 
$K^0_L\pip\pim$ vs CP-odd &  $K^0_S\piz$, $K^0_S\eta(\gamma\gamma)$\\
$K^0_L\pip\pim$ vs Flavour & $\Kp\pim$, $\Kp\pim\piz$, $\Kp\pim\pip\pim$\\ 
\hline
\end{tabular}}}
\end{center}
\caption{Selected tags for $K^0\pip\pim$ (self-conjugate modes are implied).}
\label{tab:tag1}
\end{table}

\begin{table}
{\scriptsize{
\begin{tabular}{ll}
\hline
Tag Group & Opposite-side Tags \\ \hline
$K^0_S\Kp\Km$ vs CP-even & $\Kp\Km$, $\pip\pim$, $K^0_S\piz\piz$, $K^0_L\piz$, $K^0_L\eta(\gamma\gamma)$, $K^0_L\omega(\pip\pim\piz)$, $K^0_L\eta(\pip\pim\piz)$, $K^0_L\eta'(\pip\pim\eta)$\\
$K^0_S\Kp\Km$ vs CP-odd & $K^0_S\piz$, $K^0_S\eta(\gamma\gamma)$, $K^0_S\omega(\pip\pim\piz)$, $K^0_S\eta(\pip\pim\piz)$, $K^0_S\eta'(\pip\pim\eta)$,
$K^0_L\piz\piz$\\
$K^0_S\Kp\Km$ vs $K^0 h^+ h^-$ & $K^0_S\Kp\Km$,  $K^0_L\Kp\Km$, $K^0_S\pip\pim$, $K^0_L\pip\pim$\\      
$K^0_S\Kp\Km$ vs Flavour & $\Kp\pim$, $\Kp\pim\piz$\\
$K^0_L\Kp\Km$ vs CP-even & $\Kp\Km$, $\pip\pim$, $K^0_S\piz\piz$\\
$K^0_L\Kp\Km$ vs CP-odd & $K^0_S\piz$, $K^0_S\eta(\gamma\gamma)$, $K^0_S\omega(\pip\pim\piz)$, $K^0_S\eta(\pip\pim\piz)$, $K^0_S\eta'(\pip\pim\eta)$\\
%$K^0_L\Kp\Km$ vs $K^0 h^+ h^-$ & $K^0_S\Kp\Km$\\
$K^0_L\Kp\Km$ vs Flavour & $\Kp\pim$, $\Kp\pim\piz$\\
\hline
\end{tabular}}}
\caption{Selected tags for $K^0\Kp\Km$ (self-conjugate modes are implied).}
\label{tab:tag2}
\end{table}

%Kinematics and topological criteria are used to select \Kp, \pip, \piz, %$\omega$, and $\eta$ mesons. 
%Combinatorial background are estimated from sidebands and peaking background %from Monte Carlo. 
Background levels vary from 1 to 10\% of the signal
for $D\to K^0_{S,L}\pip\pim$, and from 5 to 30\% for $D\to K^0_{S,L} \Kp\Km$. In the latter case, larger background values are expected because the branching fraction is approximately six times smaller. In addition, charged kaons from $D\to K^0_{S,L}\Kp\Km$ have a soft momentum spectrum and may decay in flight. In order to mitigate these problems, additional $D$-tagging modes are reconstructed, and looser track-quality cuts are used to select charged kaons.

\section{Results for $D\rightarrow\KS\pip\pim$ and $D\rightarrow\KS\Kp\Km$}

\begin{figure}[!hbpt]
\begin{center}
\includegraphics[height=40mm]{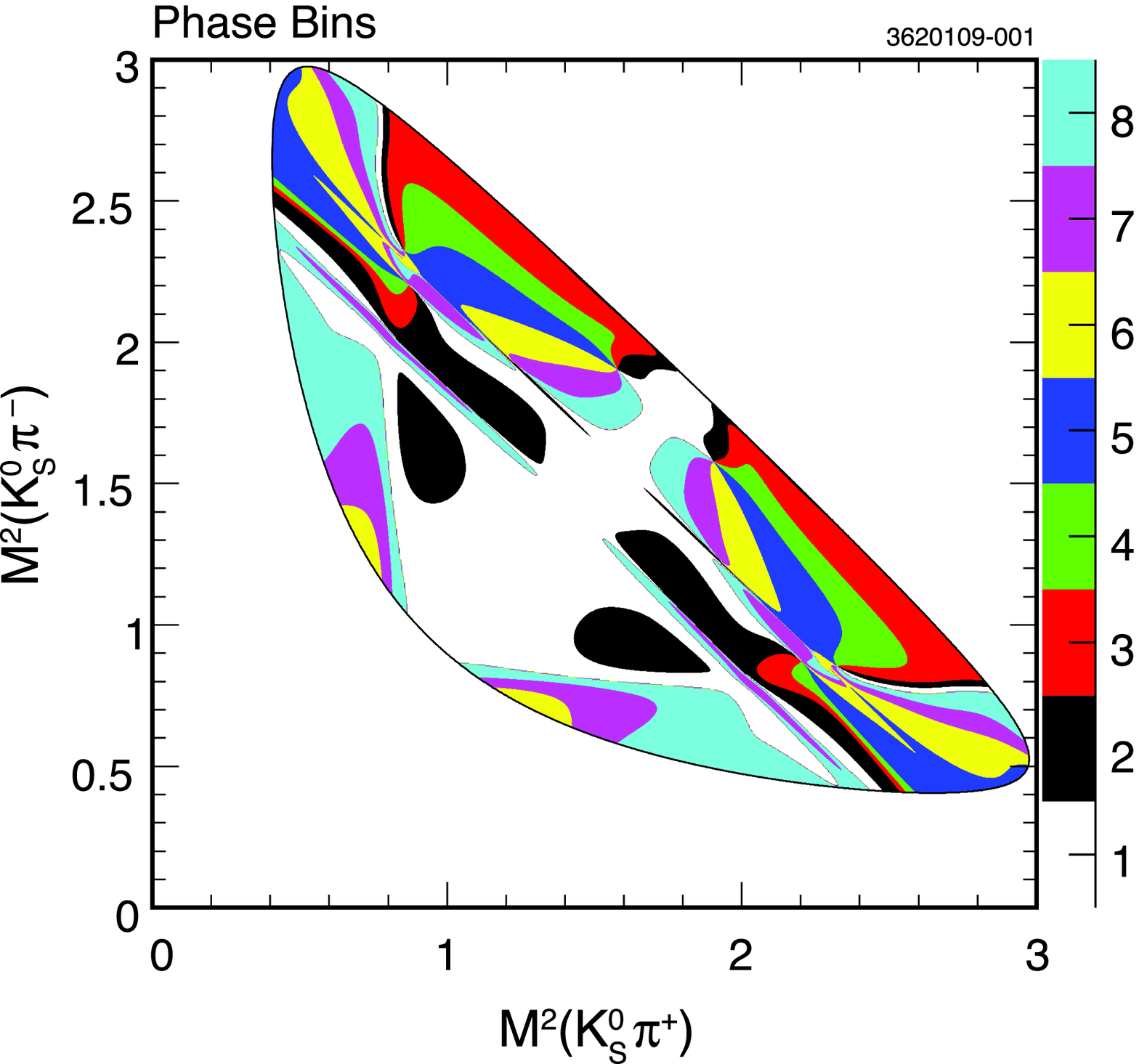}
\includegraphics[height=40mm]{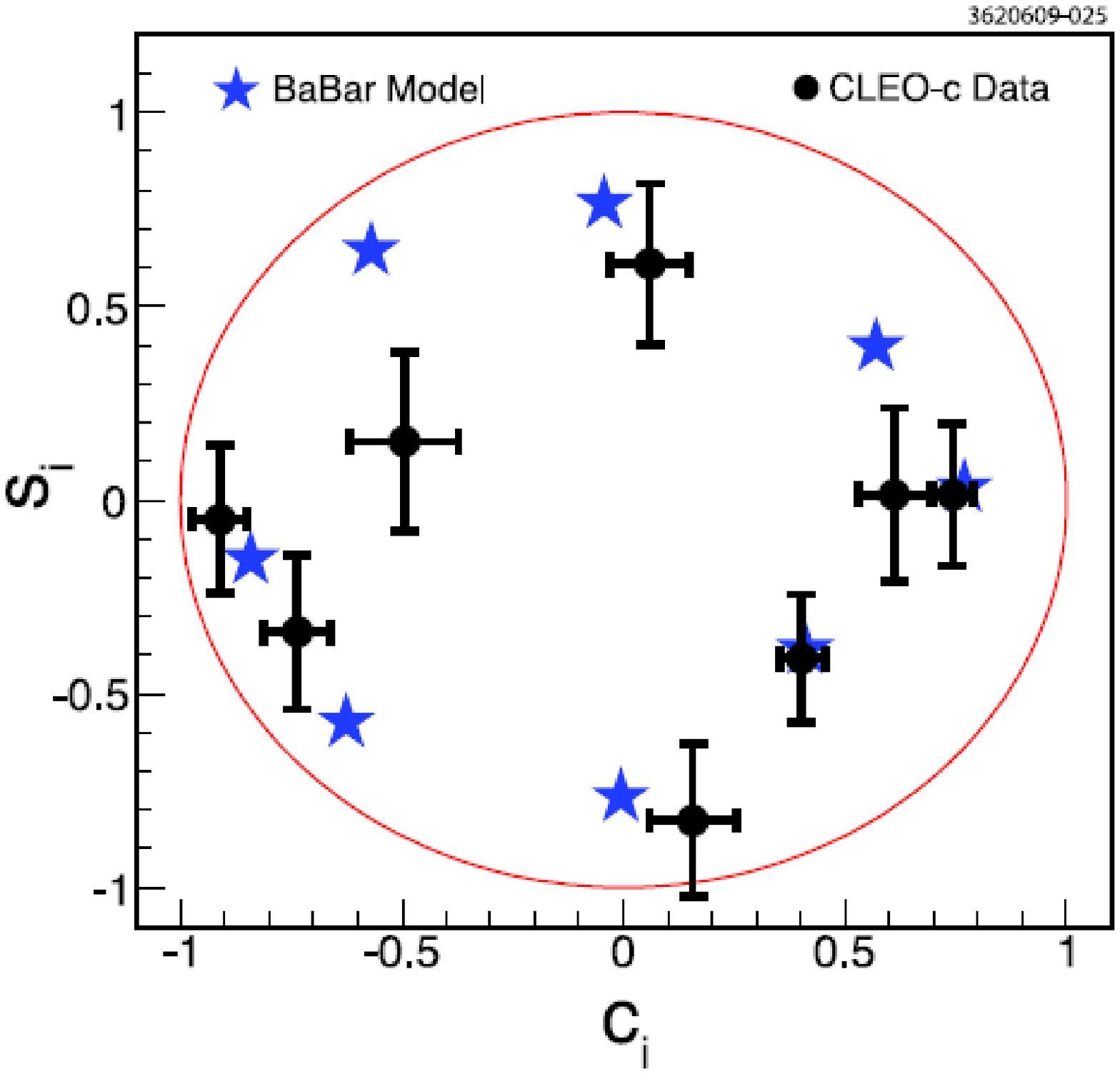}
\end{center}
\caption{$D\to \KS\pip\pim$: $\Delta\delta_D$ binning of the Dalitz plot (left); results for $c_i$ and $s_i$ (right). Error bars indicate the measured values; stars indicate the predicted values from the BaBar isobar model~\cite{isobarBaBar}.}
\label{fig1}
\end{figure}

The measured values of $c_i$ and $s_i$ for $D\rightarrow\KS\pip\pim$ are shown in Fig.~\ref{fig1}. They are in good agreement with the predicted values, computed from existing models. The systematic uncertainties, which are included in the shown error bars, are relatively small.

\begin{figure}
\begin{center}
\includegraphics[height=40mm]{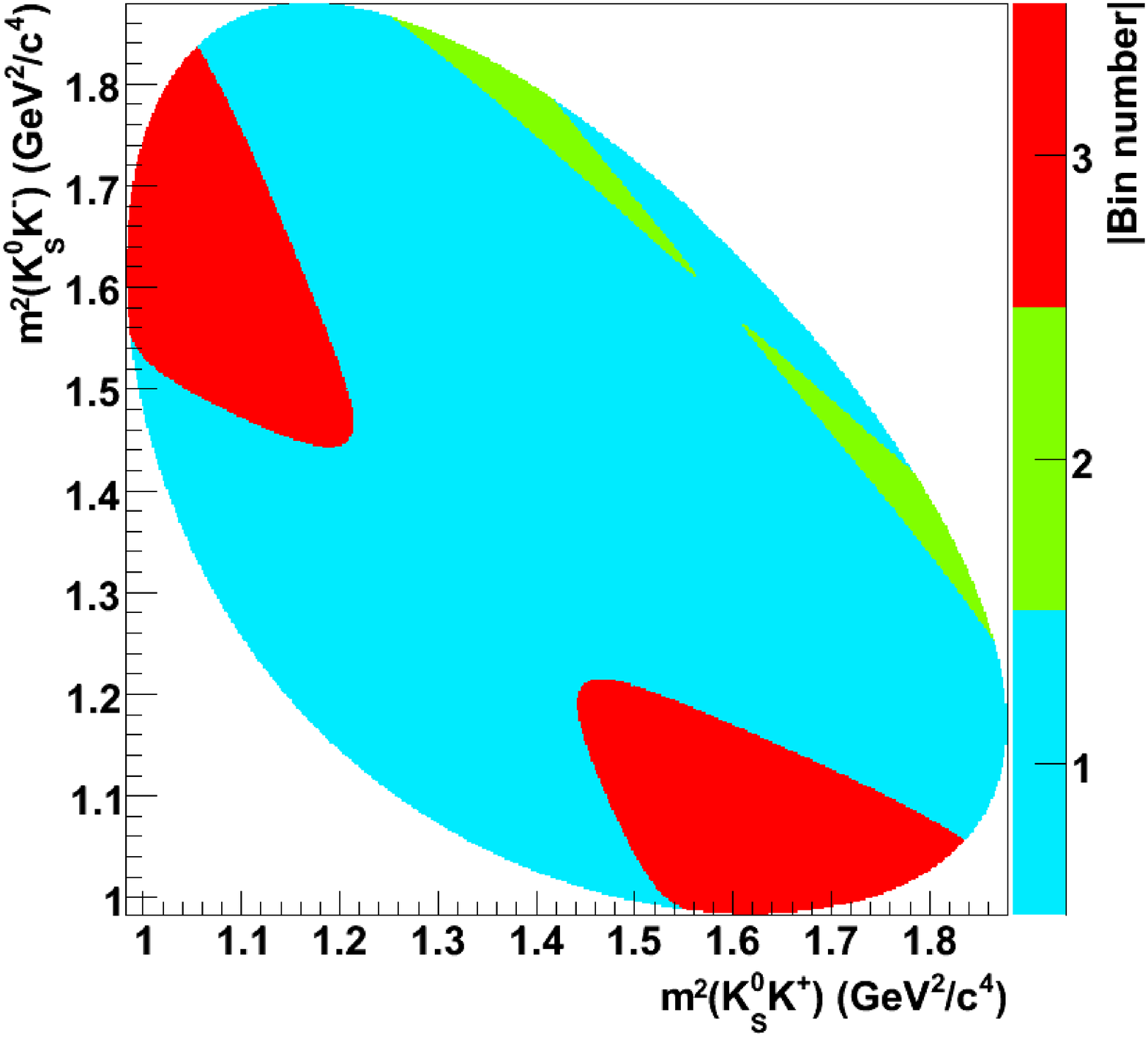}
\includegraphics[height=40mm]{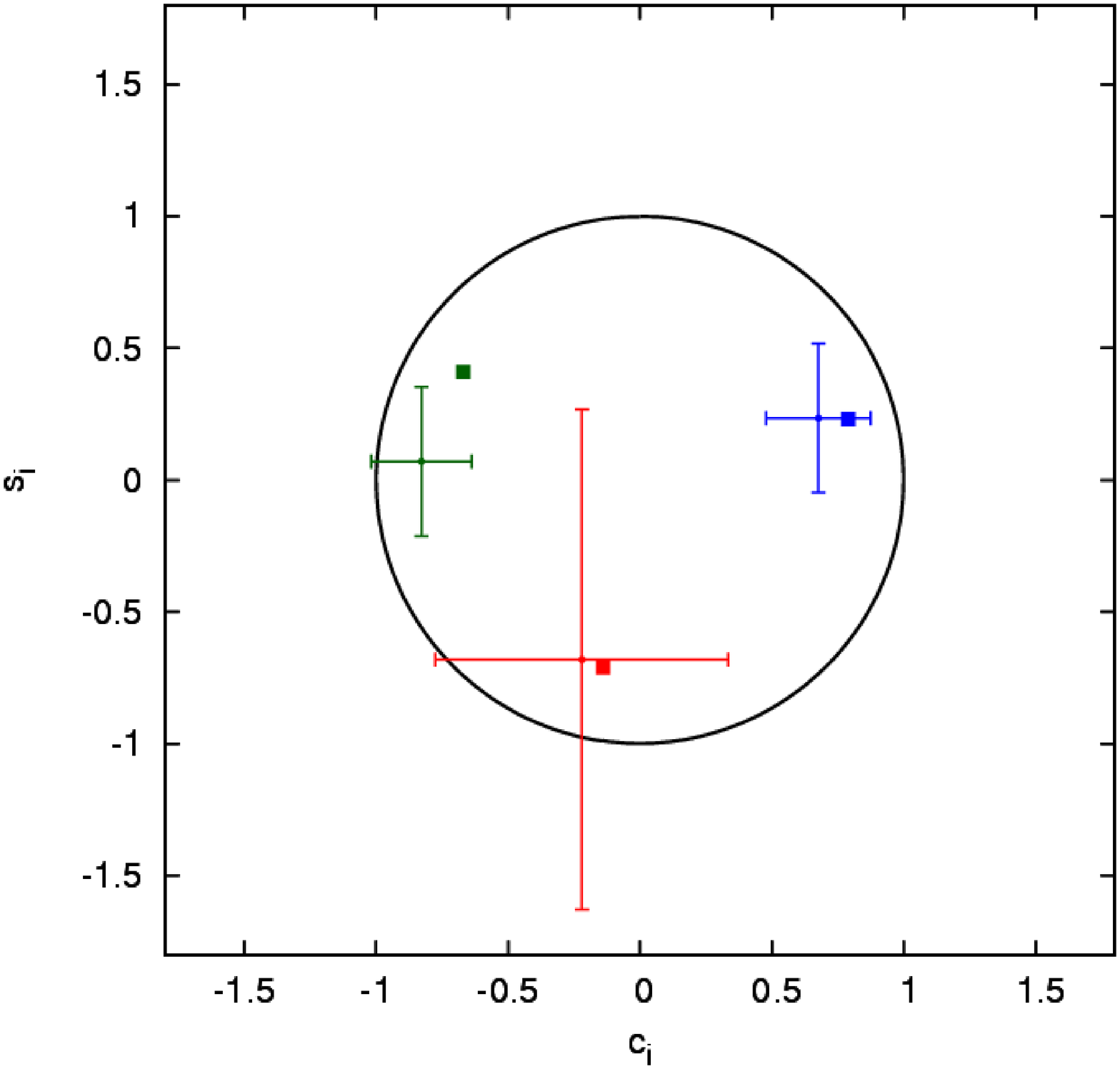}
\end{center}
\caption{$D\to \KS\Kp\Km$: $\Delta\delta_D$ binning of the Dalitz plot (left); results for $c_i$ and $s_i$ (right). Error bars indicate the preliminary CLEO-c results (statistical errors only); squared dots indicate the predicted values from the BaBar model~\cite{BaBar}.}
\label{fig2}
\end{figure}

Preliminary results for $c_i$ and $s_i$ for $D\rightarrow\KS\Kp\Km$ are shown in Fig.~\ref{fig2}. They are also in good agreement with the values predicted by the BaBar model.
In this case, the systematic uncertainties have not been evaluated yet. The statistical error is expected to dominate, since yields are small.

\section{Impact on the measurement of the CKM-angle $\gamma$}
These measurements will have an important role in the determination of the CKM angle $\gamma$ using $\Bpm\to D(\KS h^+ h^-)\Kpm$ decays. 
%In fact, neglecting CP-violation in \Dz-\Dzb mixing, the total decay amplitude %$A(\Bpm\to D(\KS h^+ h^-)\Kpm)$ is proportional to $f_D(m_+, m_-) + r_B %e^{i(\delta_B\pm\gamma)} f_D(m_-, m_+)$, where $r_B$ is the relative magnitude %of the suppressed amplitude ($\Bm\to\Dzb\Km$) to the favoured one %($\Bm\to\Dz\Km$), $\delta_B$ is their strong-phase difference, and $f_D(m_+, %m_-) = |f_D(m_+, m_-)| e^{i\delta_D(m_+,m_-)}$ is the $D$-decay amplitude. The %strong-phase difference in the $D$ decay, $\Delta\delta_D = \delta_D(m_+, m_-) %- \delta_D(m_-,m_+)$, has to be known in order to extract $\gamma$ from the %difference in the \Bpm decay rates across the Dalitz plane. 
We recall that the current measurements of $\gamma$ from these decays have used different models for the $D$ decay amplitude, and that
the associated model systematic uncertainty is $5^{\circ}-9^{\circ}$, as estimated by BaBar~\cite{BaBar} and Belle~\cite{Belle}, respectively. These large and hard-to-quantify uncertainties will limit the precision at LHCb and future B-factory experiments. 

The $c_i$ and $s_i$ parameters measured by CLEO-c will enable experiments to measure $\gamma$ with the model-independent approach introduced by Giri {\em et al.}~\cite{GirietAl}. This is based on a fit to the number of $\Bpm$ events, $N_i^\pm$, in bins of the $D\to\KS h^+ h^-$ Dalitz plot. Since
$$ N_i^\pm \propto \{K_i + r_B^2 K_{-i} + 2 r_B \sqrt{K_i K_{-i}} [c_i cos (\delta_B \pm \gamma) + s_i sin(\delta_B \pm \gamma)]\},$$
the binned fit involves only experimental observables, i.e., the already defined $c_i$, $s_i$, and $K_i$ coefficients,
and the $B$-decay parameters, $r_B$ and $\delta_B$, to be extracted from the fit together with $\gamma$.

We have evaluated the impact of the CLEO-c results on the $\gamma$ measurement with a toy Monte Carlo study, assuming $r_B = 0.1$, $\delta_B = 130^{\circ}$, and $\gamma = 60^{\circ}$. The uncertainty on $\gamma$ is reduced to about 1.7$^{\circ}$ for $\Bpm\to D(\KS\pip\pim)\Kpm$, and to about $5^{\circ}-6^{\circ}$ for $\Bpm\to D(\KS\Kp\Km)\Kpm$. These small residual errors, due mainly to the limited size of the CLEO-c data sample, will replace the corresponding model uncertainties in future measurements based on the binned approach. 
%and can be further reduced in the future by the BES-III experiment. 

\end{document}